\documentclass[12pt]{article}
\usepackage{graphicx}
\usepackage{epsfig}
\input{epsf}
\begin{document}

\title{ Expanding Universe and its manifestations beyond the
General Relativity.}

\author{ L.M. Tomilchik$^{1}$, N.G. Kembrovskaya$^{2}$\\
{\small  $^1$Institute of Physics, National Academy of Sciences
of Belarus,}\\ {\small Nezalegnasti ave. 68, 220072 Minsk, Belarus}\\
{\small $^2$ Belarusian State University,}\\
\small{Minsk, 220004, Belarus } }
\date{}
\maketitle
\begin{abstract}
It has been demonstrated that a modern stage of the Universe
expansion may be described in accordance with the observations
within the scope of the space-time conformal geometry. The clock
synchronization procedure in SR has been generalized to the case of
the expanding space. It has been found that a universal local
manifestation of the cosmological expansion is a background
acceleration, the value of which is determined by Hubble constant.
The formulae defining an explicit red-shift dependence of the
cosmological distance and expressions for Hubble law have been
obtained in a pure kinematic way from the conformal group
transformation, providing a quantitative representation of the
Pioneer anomaly and of the effect associated with the experimentally
revealed Metagalaxy transition to its accelerated expansion.
\end{abstract}

%*****************   The Body of the Article:   *************************

\section{Introduction}
The existence of the experimentally recorded local manifestations of
a cosmological expansion unexplained by a conventional gravitation
theory has received much thought since the discovery of the Pioneer
anomaly (PA). It is well known that this phenomenon - systematic
violet frequency shift detected by radio signals of the Pioneer10/11
space craft - still calls for a consistent explanation [1-4]. It is
customary to interpret the observed effect as a result of an
additional uniform acceleration $(a_p)$ of the space craft,
calculated as $a_p=(8.74\pm 1.33)10^{-8} cm/c^2$ and directed
(approximately) toward the Sun. A measurand in this case is the
propagation-time proportional frequency shift $\triangle\nu_a$ in a
signal from the transmitter located at each space craft. The
experimentally recorded uniform rate of this shift $
\triangle\dot{\nu}_a=\frac{d(\triangle\nu_a)}{dt}$ is equal to [1,2]
\begin{equation}
\label{GrindEQ__1_1_}
 \triangle\dot{{\nu}}_a=(5.99\pm
0.01)10^{-9} Hz/s.
\end{equation}
The shift direction observed is in line with a decrease in the
expected red shift (as a transmitter recedes from the observation
point), i.e. the shift is actually violet and anomalous by its
nature.

According to the conventional treatment based on a natural
supposition concerning the Doppler origin of the shift, this effect
may be explained, with regard to a nonrelativistic character of a
relative motion of the object, as follows.

A red frequency shift of the signal $\nu_{obs}$ from the transmitter
operating at the frequency $\nu_0$ and receding from the observation
point with the speed $V\ll c$ is determined by the well-known
formula from a theory of Doppler effect \footnote{ Here and
hereinafter the consideration is given to the radial speed component
and hence to the longitudinal Doppler effect.} as
\begin{equation}
\label{GrindEQ__1_2_}
\nu _{obs} =\nu _{0} \left(1-{\raise0.7ex\hbox{$ V
$}\!\mathord{\left/{\vphantom{V
c}}\right.\kern-\nulldelimiterspace}\!\lower0.7ex\hbox{$ c $}}
\right).
\end{equation}
With a negative uniform acceleration that is equal to --W\textit{a},
the expression for speed is of the form
\begin{equation} \label{GrindEQ__1_3_}
V=V_{0} -W_{{\rm a}} t,
\end{equation}
where $V_0$ -- constant speed component.
 From this we have

\begin{equation} \label{GrindEQ__1_4_}
\Delta \nu _{{\rm a}} \mathop{=}\limits^{def} \nu _{obs} -\nu _{0}
\left(1-{\raise0.7ex\hbox{$ V $}\!\mathord{\left/{\vphantom{V
c}}\right.\kern-\nulldelimiterspace}\!\lower0.7ex\hbox{$ c $}}
\right)=\nu _{0} \frac{W_{{\rm a}} }{c} t.
\end{equation}
Consequently, the rate $\dot{\nu }_{{\rm a}} $ of the anomalous
violet radiation-frequency drift equals
\begin{equation} \label{GrindEQ__1_5_}
\dot{\nu_{a}}=\nu_{0}\frac{W_a}{c}.
\end{equation}
Comparing (\ref{GrindEQ__1_5_}) with (\ref{GrindEQ__1_1_}), a
numerical value of the anomalous acceleration $W_{{\rm a}} $ may be
determined from
$$ W\cong c/\nu_0 6\cdot 10^{-9} cm/s^2, \nonumber$$

where $\nu _{0} =2.29\cdot 10^{9} Hz$ - operating frequency of the
Pioneer 10/11space craft transmitter. Hence it follows that $W\cong
7.86\cdot 10^{-8} cm/s^2 $, coincident with the above-mentioned
acceleration $a_p$ within the limits of a permissible value.

Obviously, the treatment considered necessitates the existence of
the dynamic factors really causing such acceleration. But it has
been impossible to explain adequately this effect due to the
detection of the corresponding gravitating sources (see [1-4]). As
noted in [1], all the efforts were in vain \textit{«facing a stiff
experimental wall».}

Quite natural in such a situation is to search for alternative
approaches to possible physical causes of the observed effect. One
of the approaches is associated with the assumption that there
exists some time ``inhomogeneity'' analytically represented as a
nonlinear function $t'(t)$, where t denotes ``homogeneous'' time.
Then we can introduce the notion of ``clock acceleration'' (see
[1,2]) $\alpha _{t} $ as follows:

\begin{equation} \alpha _{t} \mathop{=}\limits^{def} \frac{d^{2}
t'}{dt^{2} }  ~\mbox{$(dimension - (time)^{-1})$}.
\label{GrindEQ__1_6_}
\end{equation}
\noindent For \textit{$\alpha _{t} $} to be constant, nonlinearity
must be quadratic that, in combination with the apparent
requirement$\left(\frac{dt'}{dt} \right)_{t\to 0} \to 1$, leads to
the following function:
\begin{equation} \label{GrindEQ__1_7_}
t'\left(t\right)=t+\frac{\alpha _{t} }{2} t^{2} .
\end{equation}
As the expected nonlinearity effect should be small, we assume
fulfillment of the condition $\alpha _{t} t<<1$.

It is easily seen that such an \textit{ad hoc} hypothesis in fact
explains the observed frequency shift without having recourse to the
existence of an additional acceleration for the signal source.
Indeed, let $\Delta t$and $\Delta t'$be respectively the times of
the emitted and received signals, considered as a finite wave train
containing a particular fixed number (N) of complete oscillations.
In the process it is assumed that both time intervals $\Delta t$and
$\Delta t'$ are considerably less than the signal propagation period
(t). The frequencies of the emitted ($\nu_0 $) and received
($\nu_{obs} $) signals are determined from

\[\nu _{0} =\frac{N}{\Delta t} ,\nu _{obs} =\frac{N}{\Delta t'} .\]
Based on (\ref{GrindEQ__1_7_}) and with due regard for the
requirements $\Delta t,\Delta t'<<t$, we get

\[\Delta t'=\Delta t\left(1+\alpha _{t} t\right),\]
from where we have

\[\nu _{obs} =\nu _{0} \left(1+\alpha _{t} t\right)^{-1} =\nu _{0} \left(1-\alpha _{t} t\right).\]
From this equation for the frequency shift $\Delta \nu _{0}
\mathop{=}\limits^{def} \nu _{obs} -\nu _{0} $ we derive the formula

\[\Delta \nu _{{\rm a}} =-\nu _{0} \alpha _{t} \cdot t,\]
with which an expression for the frequency shift rate  $\dot{\nu
}_{{\rm a}} $ may be determined by

\begin{equation} \label{GrindEQ__1_8_}
\dot{\nu }_{{\rm a}} =\frac{d\left(\Delta \nu _{{\rm a}}
\right)}{dt} =-\nu _{0} \alpha _{t} .
\end{equation}
Comparing (\ref{GrindEQ__1_8_}) with the experimental value of
(\ref{GrindEQ__1_1_}), we can determine a numerical value for
$\alpha _{t} $from

\[\alpha _{t} =\nu _{0} ^{-1} \cdot 6\cdot 10^{-9} s^{-1} =\frac{6}{2. 29} \cdot 10^{-18} s^{-1} \cong 2.62\cdot 10^{-18} s^{-1} .  \]
A numerical value of the anomalous acceleration W\textit{a} is given
by $W_{{\rm a}} =c\alpha _{t} $, where c is the speed of light. It
is clear that in this approach the quantity $W_{{\rm a}} $ could not
be treated as a \textit{real} (or caused by dynamic factors)
acceleration.  It is rather a matter of its specific imitation:
acceleration mimicry of a kind.

A very intriguing fact of the closeness between the numerical value
of $\alpha _{t} $ and Hubble constant determined from the
astrophysical data has been noted by several authors, beginning from
the research group that had discovered ÐÀ (see [1,2]). And the
efforts to relate the quadratic time nonlinearity to cosmological
expansion followed almost immediately.

The fact that the actual electromagnetic-signal propagation time
$(t')$ in the expanding space should be ``shifted'' relative to the
conventionally assumed time $\left(t=\frac{1}{2} \left\{t^{fin}
-t^{in} \right\}\right)$ according to

\begin{equation} \label{GrindEQ__1_10_}
t'=t-\frac{H_{0} }{2} t^{2} ,
\end{equation}
where $H_0$ -- Hubble constant, has been first demonstrated in [5]
and supported subsequently in [6].

Actually, in [5,6] it has been found that in any Friedmann-type
model  (Robertson-Walker metric $dS^{2} =c^{2} dt^{2} -\chi ^{2}
\left(t\right)dr^{2} $) on condition $\frac{d\chi }{dt} >0$
(expanding space) an expression for the light signal propagation
time in the approximation, linear with respect to Hubble constant,
is automatically containing the quadratic nonlinearity.\footnote{
Note, however, that the treatment within the scope of GR based on
the Schwarzschild solution at the background of Friedmann's metric
results in the quadratic function of Hubble constant   (see [7]),
and this is considerably lower than the observed value. } The
approach associated with a treatment of PA as a local manifestation
of the cosmological expansion has been developed in a number of
works (e.g., see [8-10] and the references), where the situation is
treated in terms of the ``clock acceleration''.

As it has been first shown in [11], the quadratic nonlinearity of
the form (\ref{GrindEQ__1_7_}) with $\alpha _{t} =-H_{0} $ arises
naturally within the scope of the clock synchronization procedure
generalization by Einstein-Poincar\'{e} to the case of the expanding
space.  Then (see [12,13]) it has been found that this nonlinearity
originates in the assumption $H_{0}t \ll 1$ as a consequence of the
conformal time inhomogeneity. By the approach based on the notion of
time inhomogeneity, an anomalous violet electromagnetic-radiation
frequency drift in the location-type experiments is interpreted as a
universal local kinematic effect caused by the cosmological
expansion.  This effect is independent of the presence of some real
gravitating sources, and in the appropriate experimental conditions,
in principle, it should be observed at any frequencies. From this
viewpoint, PA is the first actually recorded effect of this kind.
Practical detection of the presence (or of the absence) of an
anomalous violet shift in other (e.g., optical) frequency range may
contribute much to elucidation of the PA effect physical nature. An
idea of the corresponding experiment has been put forward  in [14].

This paper presents elementary theoretical considerations to
substantiate the fact that the anomalous frequency shift
$\left(\Delta \nu _{{\rm a}} \right)$ is determined by

\begin{equation} \label{GrindEQ__1_11_}
\Delta \nu _{{\rm a}} =\nu _{0} H_{0} t,
\end{equation}
where $\nu _{0} $ -- radiation frequency of a steady source, $H_0$
-- Hubble constant, t -- electromagnetic signal propagation time,
whereas the, experimentally detected near the red shift value
$z_{\exp } =0.46\pm 0.13$, transition from a decelerated Universe
expansion to the accelerated one may be derived as a kinematic
inference from the conformal space-time geometry.

 Among the problems considered, we can name the following:

\noindent --  Relationship between the process of the Pioneer 10/11
space craft tracking and the clock synchronization procedure in SR.

\noindent -- Generalization of the clock synchronization procedure
by Einstein-Poincar\'{e} to the case of the expanding space and
quadratic time inhomogeneity.

\noindent -- The background acceleration effect and observer's
noninertial reference system (RS).

\noindent -- Derivation of an expression for Hubble law as a
kinematic inference from the conformal space-time geometry.
Consideration of the universal violet frequency shift in the
location-type experiments.

\noindent -- Nonlinear deviations from Hubble law and the
accelerated expansion effect without the dark energy.

\section{Tracking of the Pioneer 10/11 space craft and clock synchronization procedure in SR}
\setcounter{equation}{0}

 Each communication session with a space
craft enabling measurements of its flying path parameters by
observations from the Earth is organized as follows.

At the initial time $t^{in}$, set by the tracking station clock, a
radio signal is transmitted in the direction of a receding space
craft, and, being within the reach of this space craft, the signal
initiates an airborne transmitter operating at the frequency $\nu_0
$ = 2.29·109 Hz. The transmitted signal is received at the Earth in
time $t^{fin}$ by the station clock. The time interval

\begin{equation} \label{GrindEQ__2_1_}
t=\frac{1}{2} \left\{t^{fin} -t^{in} \right\}
\end{equation}
is considered as a propagation time of the radio signal from the
moment of its sending to the instant of time when the airborne
transmitter is initiated. The distance R between the space craft and
the observation point at time t indicated by the station clock may
be calculated from

\begin{equation} \label{GrindEQ__2_2_}
R=ct,
\end{equation}
where c -- signal propagation speed (i.e. speed of light). A speed
of the receding space craft at the same time is determined from the
magnitude of Doppler red shift for the operating frequency of a
satellite transmitter. Both these relations are in obvious agreement
with the laws of electrodynamics and hence with the SR concept. In
this manner we can find numerical values for two basic
characteristics (position and speed) of the trajectory for each of
the space crafts. On the other hand, it is well known that by SR the
determination of synchronism between two spatially separated events
in each inertial reference system (IRS) specified is based on the
use of the idealized time synchronization procedure. This procedure
includes the following basic elements:

\noindent -- transmission of an instantaneous electromagnetic signal
from the point A, at the instant of time $t_{A}^{\circ } $ by the
clock localized at this point, in a direction of the point B;

\noindent -- arrival of the signal in the point B at time $t_B$ by
the clock located at this point, and its instantaneous
retransmission in the direction of the point  A;

\noindent -- return of the signal to the point A in time $t_A$ by
the clock located at this point.

\noindent In this way we have the time interval

\begin{equation} \label{GrindEQ__2_3_}
\Delta t_{A} =t_{A} -t_{A}^{\circ }
\end{equation}
measured by the clock À.

Next the clock B indications are set by two measured indications
$t_{A}^{\circ } $ and $ t_A$ of the clock À so that \textit{by
definition} its going be synchronized with the clock À in IRS
specified. To this end, we \textit{postulate} the coincidence
requirement for the time intervals of the synchronizing signal
propagation in the forward $\left(t_{AB} =t_{B} -t_{A}^{\circ }
\right)$ and backward $\left(t_{BA} =t_{A} -t_{B} \right)$
directions. From this we get

\begin{equation} \label{GrindEQ__2_4_}
t_{B} -t_{A}^{\circ } =t_{A} -t_{B}
\end{equation}
representing, from the mathematical viewpoint, a linear equation for
$t_B$. Its solution is of the form

\begin{equation} \label{GrindEQ__2_5_}
t_{B} =\frac{1}{2} \left(t_{A} +t_{A}^{\circ } \right).
\end{equation}
It is obvious that we obtain

\[t_{BA} =t_{AB} =\frac{1}{2} \left(t_{A} -t_{A}^{\circ } \right).\]

It is easily seen that the set of operations with the help of which
a distance to the Pioneer 10/11 space craft has been determined in
the process of each communication session actually repeats the
features of the idealized time synchronization procedure in SR.
Indeed, after obvious identification of the symbols in
(\ref{GrindEQ__2_1_}) and (\ref{GrindEQ__2_3_}) --
(\ref{GrindEQ__2_5_})

\[t^{in} =t_{A}^{\circ } ,\quad t^{fin} =t_{A} ,\]
it is clear that the time interval $t=\frac{1}{2} \left(t^{fin}
-t^{in} \right)$, in terms of which we calculate a distance to the
space craft, in fact may be determined as

\begin{equation} \label{GrindEQ__2_6_}
t=t_{B} -t_{A}^{\circ } =t_{A} -t_{B} ,
\end{equation}
where $t_{A} =t^{fin} $ -- time of the signal arrival in the
observation point, $t_B$ -- time of a signal emission by the space
craft, determined by (\ref{GrindEQ__2_5_}).

s regards the synchronizing signal speed c involved in formula
(\ref{GrindEQ__2_2_}) for the distance, according to Einstein  (see
[15]), it is initially defined as a ratio between the total distance
covered by the signal in the forward $(R_{AB})$ and backward
$(R_{BA})$ directions and the time interval $t_{A} -t_{A}^{\circ }
$measured by the clock À (average speed):

\begin{equation} \label{GrindEQ__2_7_}
c=\frac{R_{AB} +R_{BA} }{t_{A} -t_{A}^{\circ } } .
\end{equation}
Making an assumption, natural for the pair of the reciprocally
stationary clocks, that $R_{AB}$ and $R_{BA}$ are coincident, we get

\[c=\frac{2R_{AB} }{t_{A} -t_{A}^{\circ } } ,\]
from where with regard to (\ref{GrindEQ__2_5_}) we obtain

\begin{equation} \label{GrindEQ__2_8_}
R_{AB} =\left|R_{A} -R_{B} \right|=c\left(t_{A} -t_{B} \right).
\end{equation}
By raising (\ref{GrindEQ__2_8_}) to the square, we arrive at the
fundamental light similar interval of SR (light cone equation):

\begin{equation} \label{GrindEQ__2_9_}
S_{AB}^{2} =c^{2} \left(t_{A} -t_{B} \right)^{2} -\left|R_{A} -R_{B}
\right|^{2} =0.
\end{equation}

Considering (\ref{GrindEQ__2_6_}), it is clear that expression
(\ref{GrindEQ__2_2_}) used in calculations of a distance from the
observation point to the space craft is practically coincident with
formula (\ref{GrindEQ__2_8_}) that, within the scope of SR,
determines a distance covered by the signal in the \textit{positive}
generatrix direction of the light cone with a vertex at the
reference point of the observer (point À). In accordance with the
requirement of (\ref{GrindEQ__2_4_}), this distance is coincident
with that covered by the signal in the \textit{negative} direction
of the generatrix for the light cone having its vertex at the point
B.

The foregoing description of the time synchronization procedure
repeats the first paragraph of Section I in the fundamental work of
Einstein [15].

\section{Clock synchronization in expanding space. Quadratic nonlinearity of
time and background acceleration effect. Noninertial  reference
system. }
\setcounter{equation}{0}

 Now we consider the situation
when spatial scales are varying with time. Taking into account the
structural correspondence between the real location-type experiment
and Einstein's clock synchronization procedure, let us find out what
changes are involved into this procedure by the spatial expansion.
To this end, an expression for the expansion law is written in the
following form:

\begin{equation} \label{GrindEQ__3_1_}
R^{-1} \left(t\right)\frac{dR\left(t\right)}{dt} =H_{0} ,
\end{equation}
where $R(t)$ -- time-dependent Euclidean distance, $H_0$ --
dimensional numerical constant $(time)^{-1}$.

After an elementary integration of equation (\ref{GrindEQ__3_1_}),
for the function R(t) we have

\begin{equation} \label{GrindEQ__3_2_}
R\left(t\right)=R\left(t_{0} \right)\exp \left\{H_{0} \left(t-t_{0}
\right)\right\},
\end{equation}
where $t_0$ --  arbitrary but fixed instant of time. From this
formula the expressions for distances covered by a signal in the
forward ($R_{AB}$) and backward ($R_{BA}$) directions may be written
as

\begin{equation} \label{GrindEQ__3_3_}
\left. \begin{array}{l} {R_{AB} =R\left(t_{A}^{\circ } \right)\exp
\left\{H_{0} \left(t_{B} -t_{A}^{\circ }
\right)\right\}=R\left(t_{B} \right),} \\ {R_{BA} =R\left(t_{B}
\right)\exp \left\{H_{0} \left(t_{A} -t_{B} \right)\right\}=R_{AB}
\exp \left\{H_{0} \left(t_{A} -t_{B} \right)\right\}\quad }
\end{array}\right\} .
\end{equation}

Then it is seen that

\begin{equation} \label{GrindEQ__3_4_}
\frac{R_{BA} }{R_{AB} } =\exp \left\{H_{0} \left(t_{A} -t_{B}
\right)\right\}>1,
\end{equation}
i.e. the distances $R_{AB}$ and $R_{BA}$ may be coincident in the
absence of expansion only (or for $H_0=0$). On the assumption that
the speed of light is independent of the source's rate of motion and
of the signal propagation direction, the location distances covered
by the signal in the forward and backward directions may be
determined from

\begin{equation} \label{GrindEQ__3_5_}
R_{AB} =c\left(t_{B} -t_{A}^{\circ } \right),\quad R_{BA}
=c\left(t_{A} -t_{B} \right).
\end{equation}
By substitution of (\ref{GrindEQ__3_5_}) into (\ref{GrindEQ__3_4_})
we arrive at

\begin{equation} \label{GrindEQ__3_6_}
t_{A} -t_{B} =\left(t_{B} -t_{A}^{\circ } \right)\exp \left\{H_{0}
\left(t_{A} -t_{B} \right)\right\}.
\end{equation}
As seen, the signal propagation time in the backward direction is
always in excess of that in the forward direction. Besides,
expression (\ref{GrindEQ__3_6_}) representing the rate synchronism
requirement for clocks À and  is now a transcendental equation for
$t_B$, as opposed to the previous case when the corresponding
equation (\ref{GrindEQ__2_4_}) was linear.

Moreover, the equation includes the constant $H_0$ with the backward
time dimension that determines the expansion rate.   It is clear
that the presence of such a constant sets a certain time scale
$t_{lim} \sim H_{0}^{-1} $.  Because of this, the location procedure
of clock synchronization leads to the results rather close to the
stationary case in SR but on the assumption of the constant $H_0$
smallness, i.e. when the signal propagation time $\Delta $t in both
directions meets the condition$\Delta tH_{0} <<1$. Also, it should
be noted that in the fundamental work of Einstein [15] the
derivation of Lorentz transformations form the synchronization
requirement (\ref{GrindEQ__2_5_}) has been based on considerations
concerning differentially small distances between the clocks.

So in expression (\ref{GrindEQ__3_6_}) we limit ourselves to the
terms linear  with respect to $H_{0} \left(t_{A} -t_{B} \right)$ in
the exponential expansion.  Then we have

\begin{equation} \label{GrindEQ__3_7_}
t_{A} -t_{B} =\left(t_{B} -t_{A} ^{\circ } \right)\; \left\{1+H_{0}
\left(t_{A} -t_{B} \right)\right\}
\end{equation}
representing the following quadratic equation for $t_B$:

\begin{equation} \label{GrindEQ__3_8_}
t_{B}^{2} -\frac{2}{H_{0} } \left\{1+\frac{H_{0} }{2} \left(t_{A}
+t_{A}^{\circ } \right)\right\}t_{B} +\frac{1}{H_{0} } \left(t_{A}
+t_{A}^{\circ } \right)+t_{A} t_{A}^{\circ } =0.
\end{equation}
Its roots are of the form

\[t_{B}^{\left(\pm \right)} =\frac{1}{H_{0} } +\frac{t_{A} +t_{A}^{\circ } }{2} \pm \frac{1}{H_{0} } \left\{1+\frac{H_{0}^{2} }{4} \left(t_{A} -t_{A}^{\circ } \right)^{2} \right\}^{\frac{1}{2} } .\]
Obviously, the solution, linear for $H_0$, that could be transformed
to $t_{B} =\frac{1}{2} \left(t_{A}^{\circ } +t_{A} \right)$ at
$H_{0} \to 0$ is associated with a lower sign.

In this way the expression for $t_B$ determined from the clock
synchronization location as an explicit function of the initial and
final indications of the ``reference'' clock À is of the form

\begin{equation} \label{GrindEQ__3_9_}
t_{B} =\frac{1}{2} \left(t_{A}^{\circ } +t_{A} \right)-\frac{H_{0}
}{8} \left(t_{A} -t_{A}^{\circ } \right)^{2} .
\end{equation}
Next, we determine the signal propagation time in the forward and
backward directions as

\begin{equation} \label{GrindEQ__3_10_}
t_{AB} =t-\frac{H_{0} }{2} t^{2} ,
\end{equation}

\begin{equation} \label{GrindEQ__3_11_}
t_{BA} =t+\frac{H_{0} }{2} t^{2} ,
\end{equation}
where $t=\frac{1}{2} \left(t_{A} -t_{A}^{\circ } \right)$ --
propagation time in the absence of expansion.

As follows from (\ref{GrindEQ__3_10_}), in fact, a time interval
needed for a signal to reach the synchronized clock is less than
$t=\frac{1}{2} \left(t_{A} -t_{A}^{\circ } \right)$ by $\delta
t=\frac{H_{0} }{2} t^{2} ,$ whereas expression
(\ref{GrindEQ__3_10_}) \textit{per se} is coincident with relation
(\ref{GrindEQ__1_10_}), the use of which, as we have seen in point
1, leads to the practically observable anomalous violet shift.

Thus, the expansion violates the fundamental synchronism requirement
for the spatially distant clocks (i.e. synchronism of the spatially
separated events) assumed in SR. An analytical relation determining
the indication of the clocks synchronized by two indications of the
``reference'' clock, on retention of the light-speed constancy
postulate, in the general case turns out to be nonlinear, containing
in the explicit form a constant of dimension -- Hubble constant. In
the approximation $H_{0} t<<1$ the corresponding nonlinearity is
quadratic, directly leading to formula (\ref{GrindEQ__1_11_}) for
the anomalous frequency shift. By this means the quadratic
inhomogeneity of time specially postulated as a possible source of
PA (see [1,2]) proves to be a natural corollary from generalization
of the basic clock synchronization procedure in SR for the expanding
space.  Note that in this case a quantitative (sign including)
simulation of the directly observable effect (frequency shift) is
attained using no indication of the existent acceleration of the
signal transmitter.

If on the basis of (\ref{GrindEQ__3_10_}) and (\ref{GrindEQ__3_11_})
we return to definition (\ref{GrindEQ__3_5_}) for the distances
covered by the signal in the forward ($R_{AB}$) and backward
($R_{BA}$) directions, the following expressions may be obtained

\begin{equation} \label{GrindEQ__3_12_}
R_{AB} =R_{0} -\frac{Wt^{2} }{2} ,\quad R_{BA} =R_{0} +\frac{Wt^{2}
}{2} ,
\end{equation}
where $R_{0} =ct,\quad t=\frac{1}{2} \left(t_{A} -t_{A}^{\circ }
\right)$.

\begin{equation}
W=cH_0     \label{GrindEQ__3_13_}
\end{equation}

Proceeding from the Galilean-Newtonian kinematics, the situation
looks as though in the process of synchronization the clock B were
displaced relative to the clock A at a uniform acceleration equal
$cH_0$ in the direction of point À. Because of this, we can obtain
an expression for the anomalous frequency shift of the form given in
(\ref{GrindEQ__1_4_}), from which we can get formula
(\ref{GrindEQ__1_11_}) by substitution of $W=cH_0$ from
(\ref{GrindEQ__3_13_}).  It is important that the ``acceleration''
thus obtained is not an acceleration in a dynamic sense, i.e. it has
no relation to the existence of some real sources of force effects
on the associated test bodies.

At the same time, involvement of the parameter $W=cH_0$ having the
dimension of acceleration in formulae (\ref{GrindEQ__3_12_}) and
(\ref{GrindEQ__3_13_}) for distances points to the fact that the
``extended'' reference system, where the clocks are synchronized, is
noninertial. Whereas a uniform ``acceleration''
(\ref{GrindEQ__3_13_}), in principle, must be recorded at every
spatial point of the observer's reference system, i.e. it must be
background in character.  From this viewpoint, the anomalous violet
frequency shift may be considered as experimentally recorded
manifestations of the reference system noninertiality for a fixed
observer.  This fact has been noted, specifically in [6], where the
corresponding mechanical analog (Foucault pendulum) has been
indicated to demonstrate noninertiality of the Earth as a reference
system.  By author's opinion, in the case under study a still better
(optical) analogy is represented by the well-known Sagnac effect:
experimental recording of a proper rotation of the Earth with the
use of Michelson interferometer.

Of course, origination of the background acceleration $W=cH_0$ may
be interpreted, proceeding from the equivalence principle, as the
presence of a constant background gravitational field. And the
observable frequency shift $\Delta $$\nu $\textit{a} may be
described in terms of the Einstein's gravitational frequency shift.
It is easily seen that in the considered case this shift is
precisely violet as, in accordance with a negative acceleration
sign, the effective gravitational potential at the observation point
is always higher than that at the point of the signal emission.

So, we can see that, in conditions of the expanding space, the use
of the clock synchronization location procedure by
Einstein-Poincar\'{e} that forms the basis for SR actually results
in the quadratic inhomogeneity of time, with a scale determined by
the expansion rate (numerical value of Hubble constant).

It is acknowledged that in a history of physics and in history on
the whole it is useless to speculate on how it could have been
if\dots   Nevertheless, Einstein could have received the
above-mentioned result and hence could approach the prediction of
the presence of a universal anomalous violet frequency shift should
the Universe expansion be discovered not in 1929  but as early as
the beginning of the first decade of the XX-th century when SR, as
opposed to GR, has been already framed, and what is more in two
versions.

Besides, conformal invariance of Maxwell electrodynamics established
by Batmann [16] and Kanningham [17] in 1909-1910 has opened up
possibilities for derivation of an analytical expression for Hubble
law in a pure kinematic way on the basis of space-time
transformations of the conformal group SO (4,2). In the process the
quadratic nonlinearity of time arises in the first approximation as
a consequence of its initial conformal inhomogeneity.

\section{Hubble law and violet frequency drift as a kinematic consequence of conformal space-time geometry.}
\setcounter{equation}{0}

As is known, an experimental checking of the law of cosmological
expansion calls for an independent determination of the galactic
objects recession speed and of the distances from each of these
objects to the observation point. Electromagnetic signals arriving
in the observation point from cosmologically distant sources provide
most important information.

The whole totality of modern astrophysical data bears witness to a
vanishingly small spatial curvature of the observable Universe and
to an approximate constancy of Hubble parameter, at least in the
interval of values up to the red shift on the order of unity. In
these conditions a non-Euclidean character of the space-time
manifold is actually exhibited in inhomogeneity of time only.
Because of this, it seems natural to attempt description of the
electromagnetic signal propagation in conditions of a modern
Metagalaxy, retaining an Euclidean character of 3-D space and using
some minimal extension of the standard relativistic kinematics of SR
due to the introduction of a group of special conformal
transformations changing the space-time scales.

Within the scope of this approach, an expression for Hubble law in
the assumption of the Hubble parameter constancy $H_{0}
=H\left(t_{today} \right)=const$ may be written as

\begin{equation} \label{GrindEQ__4_1_}
{\raise0.7ex\hbox{$ V $}\!\mathord{\left/{\vphantom{V R_{L}
}}\right.\kern-\nulldelimiterspace}\!\lower0.7ex\hbox{$ R_{L}  $}}
=H_{0} .
\end{equation}
Here V -- relative speed of the radiation source and detector, and
$R_L$ is determined as a distance covered by the signal arriving to
the observation point from the source emitting this signal at some
instance of time tin in the distant past. A problem of finding the
distance $R_L$ may be solved if we find the signal propagation time,
i.e. a time interval between the initial tin and final $t^{fin}$
points of the light cone generatrix in the observer's reference
system.  In this case $R_{L} \mathop{=}\limits^{def} ct$, where c --
speed of light, $t = t^{fin}- t^{in}$. In other words, $R_L$
represents what is known as location (radar) distance.

To compare formula (\ref{GrindEQ__4_1_}) with experimental data, one
needs theoretical relations to express each of the quantities V and
$R_L$ as an explicit function of one and the same set of
experimentally recorded radiation parameters.  It is well known that
a frequency shift is the most accurately measured quantity of this
kind.

We know that, within the scope of SR, a relationship between the
frequency shift

\begin{equation} \label{GrindEQ__4_2_}
z=\frac{\lambda _{obs} -\lambda _{em} }{\lambda _{em} }
\end{equation}
and relative speed (Doppler relativistic effect) is determined as a
direct kinematic consequence of Lorentz transformations.

Let us demonstrate that the use of a group of special conformal
transformations (SCT) representing, along with a Lorentz group (LG),
a subgroup of the SO(4,2) conformal group  for transformations of
the 4-D pseudo-Euclidean space-time leads to the functional
dependence of the electromagnetic signal propagation time on the
frequency shift, in turn making it possible to express the distance
$R_L$ as an explicit function of the red shift z.

The metric quadratic form $S^2=x^{\mu} x_{\mu} (\eta_{\mu\nu}  =
\textit{diag}{\{1,-1,-1,-1\}})$ of Minkowski space is invariant with
respect to Lorentz transformations (LT)

\[x'^{\mu } =L^{\mu } _{\alpha } x^{\alpha } ,L^{\mu } _{\alpha } L^{\alpha } _{\nu } =\delta ^{\mu } _{\nu } ,\]
meeting the condition

\begin{equation} \label{GrindEQ__4_3_}
x'^{\mu } x'_{\mu } =x^{\mu } x_{\mu } .
\end{equation}
Nevertheless, the Lorentz invariant ${\rm S}^{2} =x^{\mu } x_{\mu }
$ is not a scaling invariant. With respect to the transformations of
SCT group

\begin{equation} \label{GrindEQ__4_4_}
x'^{\mu } =\sigma ^{-1} \left(a,x\right)\left\{x^{\mu } +a^{\mu }
\left(x^{\alpha } x_{\alpha } \right)\right\},
\end{equation}
where

\begin{equation} \label{GrindEQ__4_5_}
\sigma \left(a,x\right)=1+2\left(a^{\alpha } x_{\alpha }
\right)+\left(a^{\alpha } a_{\alpha } \right)\left(x^{\beta }
x_{\beta } \right),
\end{equation}
$a^{\mu}$ -- four-vector parameter, it behaves as follows:

\begin{equation} \label{GrindEQ__4_6_}
x'^{\mu } x'_{\mu } =\sigma ^{-1} \left(a,x\right)x^{\mu } x_{\mu }
.
\end{equation}
But as seen from (\ref{GrindEQ__4_3_}) and (\ref{GrindEQ__4_6_}),
the requirements $x'^{\mu } x'_{\mu } =0$ and $x^{\mu } x_{\mu } =0$
in both cases are intercorrelated. This is a well-known fact of the
light-cone equation invariance with respect to LT as well as SCT.
The light-cone generating lines in both cases are, however,
subjected to transformations, that should lead to variation of the
observable frequency characteristics of electromagnetic signals on
the transition between any reference system pairs associated with
the corresponding transformation of coordinates.

First we consider a familiar situation with the relativistic Doppler
effect.

For simplicity, the four-vector x$\mu$  is chosen in the form

\begin{equation} \label{GrindEQ__4_7_}
x^{\mu } =\left\{x^{0} =ct,x,0,0\right\}
\end{equation}
and we limit ourselves to consideration of a single-parameter group
of Lorentz boosts. It is clear that we refer to the longitudinal
Doppler effect. For this case Lorentz transformations are of the
form

\noindent

\begin{equation} \label{GrindEQ__4_8_}
x'=\frac{x-\beta ct}{\sqrt{1-\beta ^{2} } } ,\quad t'=\frac{t-\beta
\frac{x}{c} }{\sqrt{1-\beta ^{2} } } ,
\end{equation}
where $\beta =\frac{V}{c} $, V -- relative speed of motion for two
inertial reference systems (IRS). In the case of an electromagnetic
signal emission a light-cone equation is of the $x^{2} -c^{2} t^{2}
=0$ in nonprimed and $x'^{2} -c^{2} t'^{2} =0$ in  primed RS.
Substituting into (\ref{GrindEQ__4_8_}) the expression $x=ct$ (for
definiteness, we select the case of the signal propagation in the
positive direction of the light cone generatrix), with regard to
$x'=ct'$we have

\begin{equation} \label{GrindEQ__4_9_}
t'=t\left(\frac{1-\beta }{1+\beta } \right)^{\frac{1}{2} }
\end{equation}
determining a linear deformation of the light-cone generatrices with
Lorentz transformation. For small time increments $\Delta t$ and
$\Delta t'$, from (\ref{GrindEQ__4_9_}) we can find that

\begin{equation} \label{GrindEQ__4_10_}
\Delta t'=\Delta t\left(\frac{1-\beta }{1+\beta }
\right)^{\frac{1}{2} } .
\end{equation}

Identifying them with oscillation periods of the emitted
($T_{emitted} =\Delta t'$) and observed ($T_{observed} =\Delta t$)
signals, respectively, and meaning by V a relative speed of the
signal source and detector separation, from (\ref{GrindEQ__4_10_})
we get

\begin{equation} \label{GrindEQ__4_11_}
\frac{T_{observed} }{T_{emitted} } =\left(\frac{1+\beta }{1-\beta }
\right)^{\frac{1}{2} } .
\end{equation}

This well-known formula for the longitudinal Doppler effect is
defining the frequency red shift. Considering the standard
definition of (\ref{GrindEQ__4_2_}), from (\ref{GrindEQ__4_11_}) we
derive the familiar relativistic expression connecting a relative
speed of the radiation source and detector to a red shift value as
follows:

\begin{equation} \label{GrindEQ__4_12_}
V\left(z\right)=c\frac{\left(z+1\right)^{2} -1}{\left(z+1\right)^{2}
+1} .
\end{equation}

Now we direct our attention to SCT determined by formulae
(\ref{GrindEQ__4_4_}) and (\ref{GrindEQ__4_5_}). As previously, we
limit ourselves to treatment of the two-dimensional (1+1) version.
The four-vector $x_{\mu}$ is considered to be given by
(\ref{GrindEQ__4_7_}). According to $[$18$]$, we select a
four-vector parameter $a^{\mu } $ of the form $a^{\mu }
=\left\{0,a,0,0\right\}$, where a$=-\frac{1}{2r_{0} } $, $r_0$ --
parameter with a dimension of distance. Transformations of
(\ref{GrindEQ__4_4_}) are conveniently written in the following
noncovariant form:

\begin{equation} \label{GrindEQ__4_13_}
x'=\frac{\xi x-\eta ct}{\xi ^{2} -\eta ^{2} } , t'=\frac{t}{\xi ^{2}
-\eta ^{2} } ,
\end{equation}
where

\[\xi =1+\frac{x}{2r_{0} } , \eta =\frac{ct}{2r_{0} } ;\]

\[\sigma \left(a,x\right)\to \overline{\sigma }\left(\xi ,\eta \right)=\xi ^{2} -\eta ^{2} .\]
In the approximation $\xi$$\approx$1, $\eta$$<$$<$1 from
(\ref{GrindEQ__4_12_}) we obtain

\[x'\cong x-\frac{c^{2} t^{2} }{2r_{0} ^{} } ,t'\cong t\]
that is associated with Galilean-Newtonian limit. Because of this,
the parameter

\begin{equation} \label{GrindEQ__4_14_}
\frac{c^{2} }{r_{0} } =w
\end{equation}
actually denotes a uniform 3-D acceleration. From this it follows
that SCT (\ref{GrindEQ__4_4_}) in the approximation considered may
be treated as a transformation from (unprimed) co-moving Lorenz
reference system (RS) S to  the (primed) noninertial RS $S'$ moving
at the uniform relative acceleration w in the + x direction. Also,
note that, in the Galilean-Newtonian approximation, relations of
(\ref{GrindEQ__4_13_}) lead to the customary acceleration
transformation rule on going from noninertial frame of reference to
the inertial one

\begin{equation} \label{GrindEQ__4_15_}
\frac{d^{2} x'}{dt'^{2} } =\frac{d^{2} x}{dt^{2} } -w.
\end{equation}

It is readily seen that in RS $S'$ the fact of noncoincident light
signal propagation times in the forward and backward directions in
the location-type experiments follows directly from the
transformations of (\ref{GrindEQ__4_13_}). This is demonstrated in
Fig. 1.

%==================Figs.01 =============================
\begin{figure}[h!]
\centerline{\psfig{figure=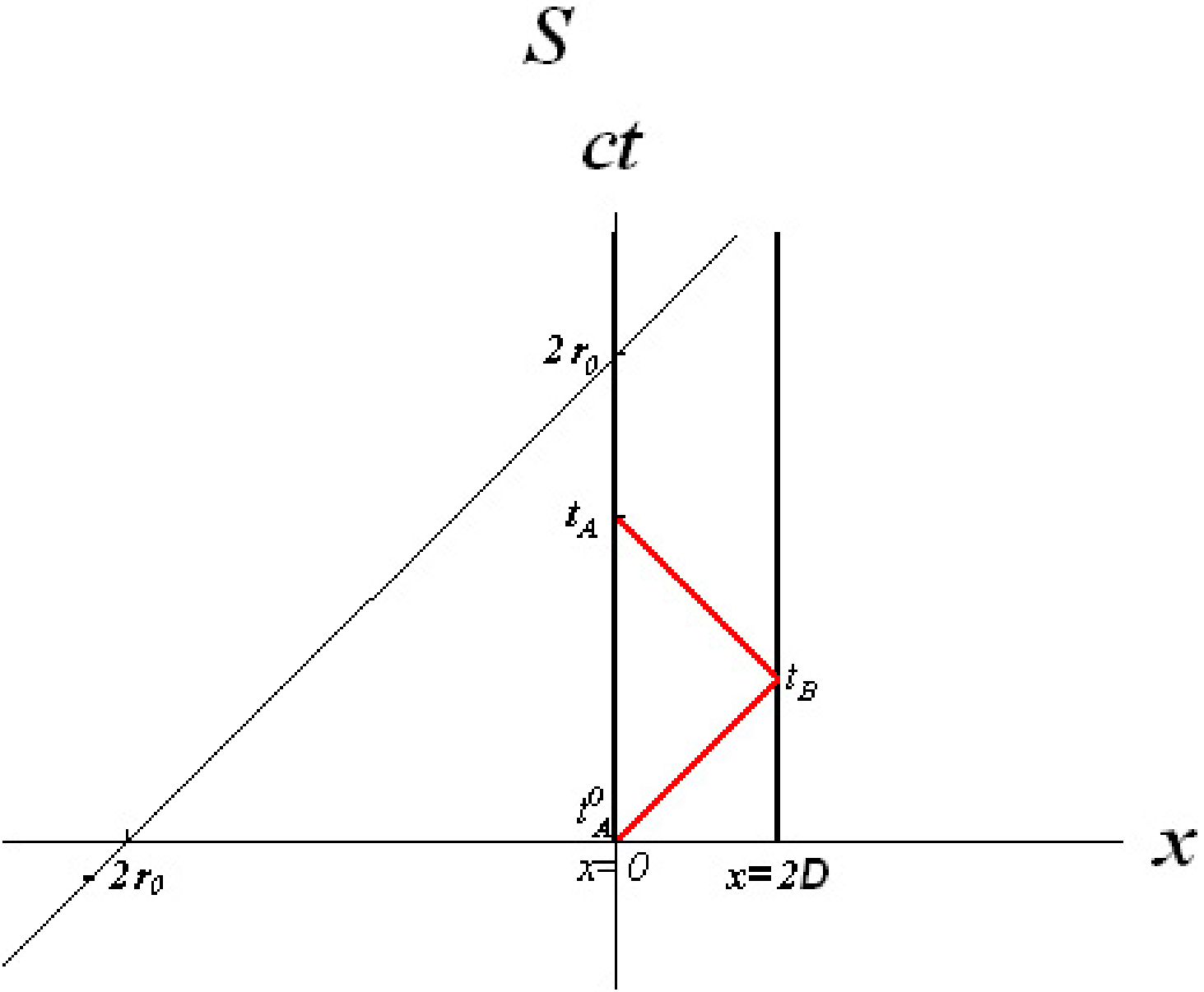,width=0.5\textwidth} }
\begin{center}
 $t_A-t_B=t_B-t^{0}_{A}$.
\end{center}
\vspace{0.5cm}
\end{figure}
%================================================
\newpage
%==================Figs.01 =============================
\begin{figure}[h!]
\centerline{\psfig{figure=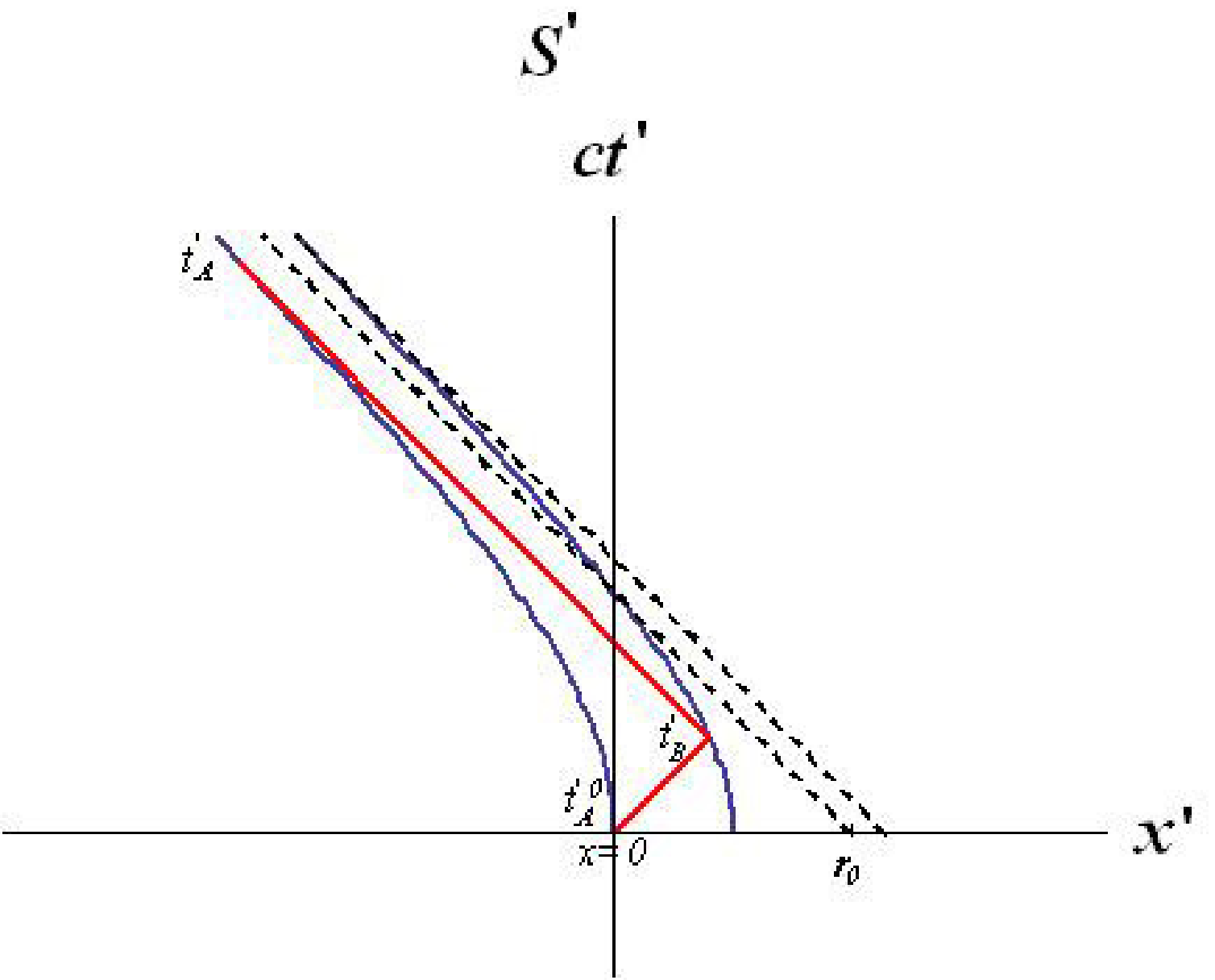,width=0.5\textwidth} }
\begin{center}
Fig 1. ${t'}_A-{t'}_B={t'}_B-{t'}^{0}_{A}$.
\end{center}
\vspace{0.5cm}
\end{figure}
%================================================

Vertical straight lines (1) and (2) on the upper denote,
respectively, observer's world line (1) and world line (2) of the
spatially distant object that is stationary relative to the origin
of RS S. As is known (see [18]), on a basis of transformation
(\ref{GrindEQ__4_13_}) these lines map into hyperbolae ($1'$) and
($2'$) in RS $S'$ (on the lower). Dotted lines represent asymptotes
of the corresponding hyperbolae.

World lines of the signal (light cone generatrices) are shown by
thin lines. At transformations (\ref{GrindEQ__4_13_}) these lines
remain straight with a slope of 45${}^\circ$ to the coordinate axes
of both reference systems.

As seen, in RS S the (segment) lengths of  generatrices are
coincident: $t_{A} -t_{B} $ and $t_{B} =t'_{A}$, whereas in RS $S'$
the length $t'_{A} -t'_{B}$ is always greater than
$t'_{B}-{t'}^{0}_{A}$.

Next we consider transformations (\ref{GrindEQ__4_13_}) for the case
of a light signal propagation, i.e. we take account of the equations
$x_{0}^{2} -x^{2} ={x'}_{0} ^{2} -{x'}^{2}=0$.

By substitution of $x=\pm ct$ into (\ref{GrindEQ__4_13_}) we can
find that

\begin{equation} \label{GrindEQ__4_16_}
x'=\pm ct\left(1\pm \frac{ct}{r_{0} } \right)^{-1} ,\quad
t'=t\left(1\pm \frac{ct}{r_{0} } \right)^{-1} ,
\end{equation}
from where it follows that SCT, retaining invariance of an equation
for the light-cone, result in the following nonlinear transformation
of its generatrices:

\begin{equation} \label{GrindEQ__4_17_}
t'_{\pm } =t\left(1\pm \frac{t}{t_{\lim } } \right)^{-1} ,
\end{equation}
where $t_{\lim } =\frac{r_{0} }{c} =\frac{c}{w} $, and signs
$\left(\pm \right)$ denote, respectively, the signal propagation in
positive and negative directions with respect to the cone vertex.
For differentially short periods $\Delta t$ and $\Delta t'$ from
\ref{GrindEQ__4_17_} we have
\begin{equation} \label{GrindEQ__4_18_}
\Delta t'_{\pm } =\Delta t\left(1\pm \frac{t}{t_{\lim } }
\right)^{-2} .
\end{equation}

Next, identifying $\Delta t$ with an oscillation period of the
emitted signal $\left(\Delta t-T_{emitted} \right)$, $\Delta t'$ --
with the period measured at the observation point $\left(\Delta
t'-T_{observed} \right)$ and assuming that the signal has been
emitted in the distant past with respect to the time of observation
(a lower sign in (4.18) should be chosen), we get the following
expression to relate wavelengths of the emitted
($\lambda_{emitted}$) and observed ($\lambda_{observed}$) signals:

\begin{equation} \label{GrindEQ__4_19_}
\lambda _{observed} =\lambda _{emitted} \left(1-\frac{t}{t_{\lim } }
\right)^{-2} .
\end{equation}

Owing to this, with the use of the standard red-shift definition\\
$\lambda _{observed} / \lambda _{emitted} =z+1$, we can find an
expression for the signal propagation time from the instant of its
emission to the time of observation in the form of the explicit
function t(z)

\begin{equation} \label{GrindEQ__4_20_}
t\left(z\right)=t_{\lim } \frac{\left(z+1\right)^{\frac{1}{2} }
-1}{\left(z+1\right)^{\frac{1}{2} } } .
\end{equation}
Therefore, the location distance $R_L$ may be determined from

\begin{equation} \label{GrindEQ__4_21_}
R_{L} =R\left(z\right)=R_{\lim } \frac{\left(z+1\right)^{\frac{1}{2}
} -1}{\left(z+1\right)^{\frac{1}{2} } } ,
\end{equation}
where $R_{\lim } =ct_{\lim } =r_{0} ={\raise0.7ex\hbox{$ c^{2}
$}\!\mathord{\left/{\vphantom{c^{2}
w}}\right.\kern-\nulldelimiterspace}\!\lower0.7ex\hbox{$ w $}} $.

\noindent To relate $R_{lim}$ to Hubble constant, we use expression
(\ref{GrindEQ__4_12_}) for V(z) and formula (\ref{GrindEQ__4_21_})
for R(z), then we write the ratio ${\raise0.7ex\hbox{$
V\left(z\right) $}\!\mathord{\left/{\vphantom{V\left(z\right)
R\left(z\right)}}\right.\kern-\nulldelimiterspace}\!\lower0.7ex\hbox{$
R\left(z\right) $}} $:

\begin{equation} \label{GrindEQ__4_22_}
 \frac{V(z)}{R(z)}=cR_{\lim } ^{-1} f(z).
\end{equation}
Here

\begin{equation} \label{GrindEQ__4_23_}
f\left(z\right)=\frac{\left(z+1\right)^{\frac{1}{2} }
}{\left(z+1\right)^{2} +1} \cdot \frac{\left(z+1\right)^{2}
-1}{\left(z+1\right)^{\frac{1}{2} } -1} .
\end{equation}
As $\mathop{\lim }\limits_{z\to 0} f\left(z\right)=2$, it may be
seen that in the limit z$<$$<$1 from (\ref{GrindEQ__4_22_}) a linear
dependence takes place in the following form:

\[V=cR_{\lim } ^{-1} R\]
coincident with the standard form (\ref{GrindEQ__4_1_}) of Hubble
law. As a result, we get the relation connecting the parameter
$R_{lim}=r_0$ to Hubble constant $H_0$ and to speed of light c

\begin{equation} \label{GrindEQ__4_24_}
R_{\lim } =2cH_{0} ^{-1} .
\end{equation}
Considering (\ref{GrindEQ__4_14_}), the acceleration w is expressed
in terms of $H_0$ and c as follows:

\begin{equation} \label{GrindEQ__4_25_}
w=\frac{1}{2} cH_{0} .
\end{equation}

As seen, an explicit analytical expression representing in the
approximation z$<$$<$1 the relationship, characteristics for Hubble
law, between the electromagnetic-signal emitter and detector
separation speed, on the one hand, and the distance covered by this
signal, on the other hand, may be derived within the scope of a
conformal group of SO(4,2).

The foregoing simplest derivation, making allowance only for the
radial motion component is exclusively based on the use of explicit
expressions (\ref{GrindEQ__4_9_}) and (\ref{GrindEQ__4_17_}) for
deformation of the light cone generatrices with respect to Lorentz
transformations and special conformal transformations. Consequently,
all the results presented in this Section are associated with pure
kinematic manifestations of the conformal space-time geometry.

It is noticeable that in this case we succeeded in finding simple
analytical expressions which incorporate the explicit red-shift
dependence of the recession speed (\ref{GrindEQ__4_12_}) as well as
of the location distance (\ref{GrindEQ__4_21_}). The function V(z)
given by (\ref{GrindEQ__4_12_}) is a well-known relativistic
expression for the longitudinal Doppler effect.  Thus, with the
derivation of relation (\ref{GrindEQ__4_12_}) given in this Section
we could aspire to a more than methodological novelty.

As regards the expression (\ref{GrindEQ__4_21_}) specifying the
cosmological location distance in the form of an explicit function
of the red shift, no other references to it in literature sources
have been found by the author. Since a numerical value of the
parameter $R_{lim}$ in (\ref{GrindEQ__4_21_}) has been determined in
accordance with (\ref{GrindEQ__4_24_}) in terms of  the familiar
physical constants, the expression for $R_L=R(z)$ may be subjected
to a direct experimental checking.  It might be advisable that the
proposed simple formula (\ref{GrindEQ__4_21_}) attracted the
attention of astrophysics specializing in the field of measurements
of the distances to the cosmologically separated objects, even
though the considerations involved look inadequately substantiated.

Note again that in the approach put forward a correlation with
Hubble law is made in the approximation of small z. Just in this
approximation  the correlation (\ref{GrindEQ__4_24_}) between the
parameter $R_{lim}$ and Hubble constant $H_0$ has been found as well
as relation (\ref{GrindEQ__4_25_}) determining the uniform
background acceleration  w in terms of the constants $H_0$ and c.

At the same time, it is seen that the approximation z$<$$<$1 is in
line with the requirement ${\raise0.7ex\hbox{$ t
$}\!\mathord{\left/{\vphantom{t t_{\lim }
}}\right.\kern-\nulldelimiterspace}\!\lower0.7ex\hbox{$ t_{\lim }
$}} <<1$ in expression (\ref{GrindEQ__4_20_}) and hence in the
initial formula (\ref{GrindEQ__4_17_}) determining the conformal
time inhomogeneity. Let us write an expression following from
(\ref{GrindEQ__4_17_}) in the approximation linear with respect to
$\frac{t}{t_{\lim } } $:

\begin{equation} \label{GrindEQ__4_26_}
t'_{\pm } =t\mp \frac{t^{2} }{t_{\lim } } .
\end{equation}
Here $t_{\lim } ={\raise0.7ex\hbox{$ c
$}\!\mathord{\left/{\vphantom{c
w}}\right.\kern-\nulldelimiterspace}\!\lower0.7ex\hbox{$ w $}} $,
and upper (lower) signs correspond to the signal propagation in the
forward (backward) directions of the light cone generatrix. Taking
account of definition (\ref{GrindEQ__4_25_}) of the parameter w,
from (\ref{GrindEQ__4_26_}) we can find

\begin{equation} \label{GrindEQ__4_27_}
t'_{\pm } =t\left(1\mp \frac{H_{0} }{2} t\right),
\end{equation}
precisely repeating relations (\ref{GrindEQ__3_10_}),
(\ref{GrindEQ__3_11_}) obtained in Section  3 within the scope of
Einstein's clock synchronization procedure generalized to the case
of expanding space. Naturally, after multiplication of both sides in
(\ref{GrindEQ__4_27_}) by c, we arrive at the formulae
(\ref{GrindEQ__3_12_}), (\ref{GrindEQ__3_13_}) determining the
distances covered by the signal in the forward and backward
directions and derived in the preceding Section, within the clock
synchronization procedure generalization to the case of the existing
expansion.

It remains only to add, as a possible historic variant for the
development of a relativistic concept, that, without doubt, the
results of this Section could have been obtained  by joint efforts
of the authors of both SR versions -- Einstein and Poincar\'{e} --
should the discovery of the Universe expansion take place, e.g., in
1911.  Indeed, the corresponding generalization of the clock
synchronization at that time, in principle, might be performed by
Einstein himself. Besides, it is well known that even in his paper
written in 1905 Einstein [15] framed a relativistic theory of
Doppler effect as a direct kinematic outcome of Lorentz
transformations.

On the other hand, such a an expert in the field of a group theory
as Poincar\'{e}, who in his pioneer works ``On a Theory of the
Electron'' [19] actually put forward a modern treatment of SR ahead
of his time by several decades, in 1911 hardly could be ignorant of
the conformal invariance of Maxwell field equations. Supposedly, for
him a solution of the problem on the use of a group of the
scaling-invariant special conformal transformations for the
description of the light signal propagation process in the expanding
space and hence on ascertaining the fact of the light-cone
generatrices conformal deformation, with the following theoretical
derivation of an expression for Hubble law and with the prediction
of anomalous violet electromagnetic-signal frequency drift in the
location-type experiments, would be a mere matter of technique.

\section{ Effect of accelerated Universe expansion without dark energy.}
\setcounter{equation}{0}

Let us consider a behavior of the function ${\raise0.7ex\hbox{$
V\left(z\right) $}\!\mathord{\left/{\vphantom{V\left(z\right)
R\left(z\right)}}\right.\kern-\nulldelimiterspace}\!\lower0.7ex\hbox{$
R\left(z\right) $}} $ over the whole interval of its application,
i.e. in the semi-infinite interval $0\le z<\infty $. Using
(\ref{GrindEQ__4_12_}) and (\ref{GrindEQ__4_21_}) for the functions
V(z) and R(z), we can write the ratio ${\raise0.7ex\hbox{$ V
$}\!\mathord{\left/{\vphantom{V
R}}\right.\kern-\nulldelimiterspace}\!\lower0.7ex\hbox{$ R $}} $ in
the following dimensionless form:

\begin{equation} \label{GrindEQ__5_1_}
\Phi \left(z\right)=\frac{V\left(z\right)}{H_{0} R\left(z\right)}
=\frac{1}{2} \frac{\left(z+1\right)^{\frac{1}{2} }
}{\left(z+1\right)^{\frac{1}{2} } +1} \cdot
\frac{\left(z+1\right)^{2} -1}{\left(z+1\right)^{\frac{1}{2} } -1} .
\end{equation}
The function $\Phi \left(z\right)$ over the interval $0\le z\le 5$
is graphically shown in Fig. 2.

%==================Figs.02 =============================
\begin{figure}[h!]
\centerline{\psfig{figure=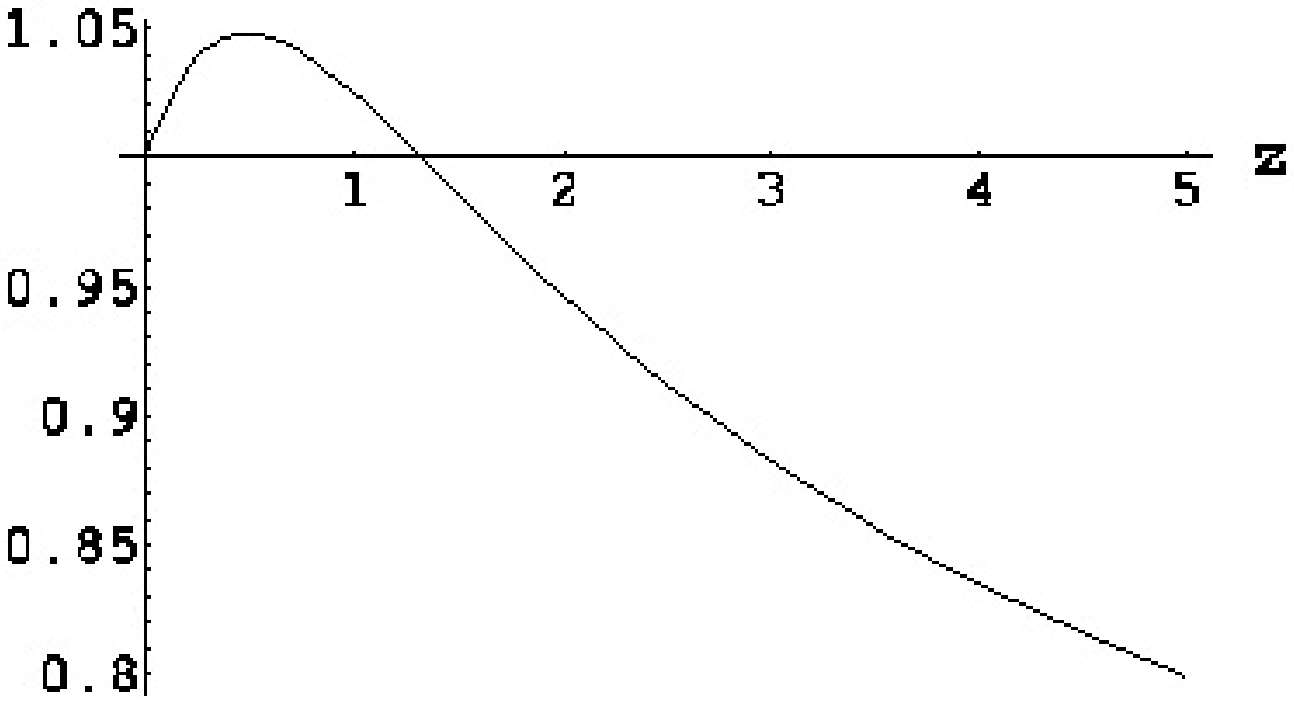,width=0.5\textwidth} } Fig 2.

\end{figure}
%================================================

As easily seen, for $z\to \infty $ the function $\Phi
\left(z\right)$ asymptotically tends to the value that is equal to
$\frac{1}{2} $. A horizontal line parallel to the z-axis is
associated with the strictly proportional function V(R). The
function $\Phi \left(z\right)$ is characterized by clear deviations
from the line $\Phi \left(z\right)$=1, which are no greater than 5\%
over the interval $0\le z\le 2$. It intersects the line $\Phi
\left(z\right)$=1 twice (at the points z=0 and $z\cong 1.315$)
having a maximum at $z_{\max } \cong 0.475$.

To have a better understanding of the reasons for such a behavior of
the function $\Phi \left(z\right)$, we represent it as $\Phi
\left(z\right)={\raise0.7ex\hbox{$ \Phi _{V} \left(z\right)
$}\!\mathord{\left/{\vphantom{\Phi _{V} \left(z\right) \Phi _{R}
\left(z\right)}}\right.\kern-\nulldelimiterspace}\!\lower0.7ex\hbox{$
\Phi _{R} \left(z\right) $}} $, where the functions

\[\Phi _{V} \left(z\right)={\raise0.7ex\hbox{$ V\left(z\right) $}\!\mathord{\left/
{\vphantom{V\left(z\right) c}}\right.\kern-\nulldelimiterspace}\!\lower0.7ex\hbox{$ c $}}
 =\frac{\left(z+1\right)^{2} -1}{\left(z+1\right)^{2} +1} \cong \left\{\begin{array}{l}
 {z\left(1-z\right)\quad for\quad {\rm z}<<{\rm 1,}} \\ {1-2z^{-2} \quad for\quad {\rm z}>>{\rm 1;}} \end{array}\right. \]

\[\Phi _{R} \left(z\right)=\frac{H_{0} R\left(z\right)}{c} =2\frac{\left(z+1\right)^{\frac{1}{2} } -1}
{\left(z+1\right)^{\frac{1}{2} } } \cong \left\{\begin{array}{l}
{z\left(1-\frac{3z}{2} \right)\quad for\quad {\rm z}<<{\rm 1,}} \\
{2\left(1-z^{-\frac{1}{2} } \right)\quad for\quad {\rm z}>>{\rm 1}}
\end{array}\right. \] determine a speed of the signal source and
detector separation and a distance covered by the signal expressed,
respectively, in units of c and $cH_{0}^{-1} $. The functions $\Phi
_{V} \left(z\right)$ (solid line) and $\Phi _{R} \left(z\right)$
(dotted line ) over the interval $0\le z<2$ are graphically shown in
Fig. 3.
%==================Figs.03 =============================
\begin{figure}[h!]
\centerline{\psfig{figure=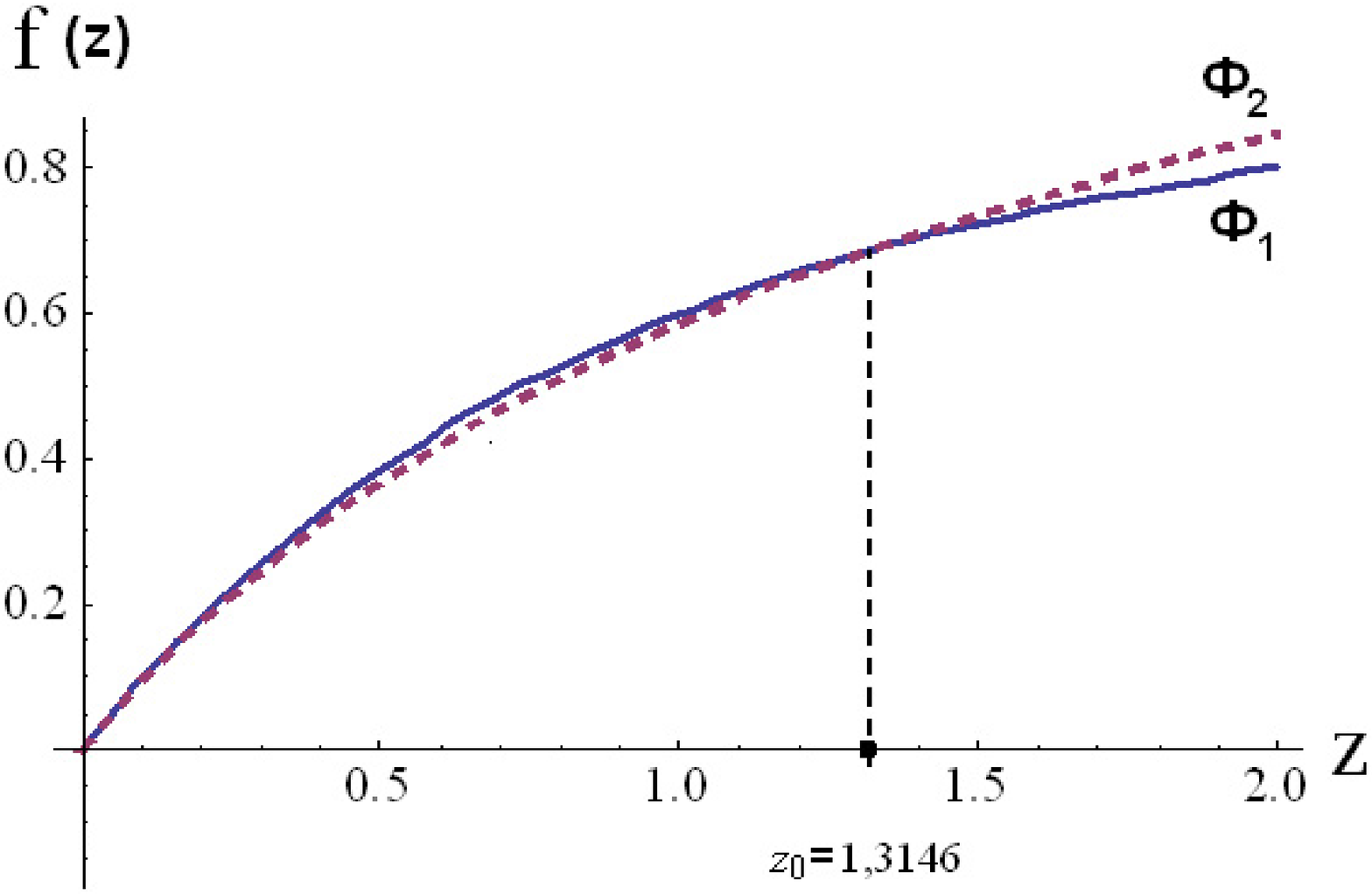,width=0.5\textwidth} }
\begin{center}
Fig 3.
\end{center}
\vspace{0.5cm}
\end{figure}
%================================================

The values of both functions are coincident at the points z=0 and
$z\cong 1.315$. Within (beyond) this interval the inequality $\Phi
_{V} \left(z\right)>\Phi _{R} \left(z\right)\quad \left(\Phi _{V}
\left(z\right)<\Phi _{R} \left(z\right)\right)$ is the case. As z is
growing from zero to $z_{\max } \cong 0.474$, the speed increases
more rapidly than the distance, the rate of increase of both
functions is equal at the point   $z_{max}$, and then the distance
is growing faster than the speed. For $z\to \infty $ the speed tends
to the limit that is equal to c, whereas the speed -- to the limit
$R_{\lim } =2cH_{0}^{-1} $. It should be recalled again that both
functional dependences $\Phi _{V} \left(z\right)$ and$\Phi _{R}
\left(z\right)$, and hence $\Phi \left(z\right)$ as well, are of a
pure kinematic origin.

Of course, the found behavioral features of the function $\Phi
\left(z\right)$ should be adequately interpreted from the viewpoint
of physics. First of all, it is required to elucidate an extent to
which the obtained theoretical dependence conforms to the data of
astronomical observations. In the case under study, we consider the
results of systematic recordings of the signals from the objects
characterized by a fairly large red shift (close to $z\approx0.5$).

It is a common knowledge that just here in 1998-1999, owing to the
advances in precise measurements of the intergalactic distances, a
sensational discovery was made.    Two independent research groups
have discovered that the distances to the corresponding
cosmologically distant sources are in fact greater than is required
by the linear Hubble law. Naturally, this fact has been interpreted
as an evidence for an increasing recession speed of galaxies, i.e.
as an indication of the accelerated Universe expansion. As
demonstrated by the experimental data, the transition from
decelerated expansion to the accelerated one has occurred in
relatively recent times. The red shift value (or the ``transition
point'') has been determined from the observations as $z_{\exp }
=0.46\pm 0.13$ (see [20,21]).

It is apparent that the expression (\ref{GrindEQ__5_1_}) derived by
us for the expansion law not only results in a linear dependence of
V(R) in the limit z$<$$<$1 but, in a surprising way, repeats the
position of  the experimentally detected extremum  (point $z_{\max }
\cong 0.475$). By our opinion, this bears witness to the fact that
formula (\ref{GrindEQ__5_1_}) describes the actually observed
Metagalaxy expansion features at the modern stage (at least,
beginning from small values of the red shift and up to
$z\longrightarrow2$). Clearly, the experimental data fitting
procedures used in this work are distinct from those adopted in
modern astrophysics for the construction of Hubble diagrams. At the
same time, as has been noted previously, a simple analytical
dependence on the red shift in our expressions containing only two
fundamental numerical parameters (c and $H_0$) is, in principle,
approachable by direct experimental checking. But, provided the
obtained values correctly simulate the observations, we still have
to find an adequate physical interpretation for deviations of the
function V(R) from linearity.

The conventional treatment is based on a cosmological model by
Friedmann--Robertson--Walker (FRW). Within its scope, the dynamic
measurand determining the Universe expansion is a deceleration
parameter that gives a change in the expansion rate. This parameter
connects the quantity and sign of the corresponding acceleration to
the physical characteristics (density and pressure) of the
gravitating matter. As this takes place, all the familiar types of
the normal matter, hypothetic nonnuclear cold dark material (CDM) as
well, contribute only to deceleration of the Universe expansion.
Because of this, the experimentally detected acceleration of this
process has lead to the knowledge that in the modern Universe there
is a special substance: peculiar vacuum-like material state
characterized by the presence of an internal tension (negative
pressure).  This hypothetical substance received the name
\textit{dark energy} or \textit{quintessence}.

According to the estimates, a share of dark energy in the modern
Metagalaxy may be as great as two thirds -- three quarters of the
total contribution made by all types of the gravitating matter. At
the same time, we have to endow quintessence with the properties not
only paradoxical from the viewpoint of the common knowledge but
practically excluding the possibility of its direct experimental
recording as a peculiar ontological substance.  It is clear that
such a situation cannot but stimulates a search for alternative
interpretations of the detected effect (see [22] and references
herein).

Now we consider a possible interpretation for the expression
(\ref{GrindEQ__4_22_}) derived by us, remaining within the scope of
the initial premises and without any special additional
suppositions.

First we should recall that in the proposed pure kinematic approach
no real sources of gravitation generating acceleration of bodies are
represented at all. An essentially new element extending the SR
kinematics is a nonlinear character of the conformal space-time
transformations. This enables one to determine an explicit
dependence between the distance and the red shift, to find an
expression for Hubble law, and to establish a key relation of the
conformal transformation parameter $t_{lim}$ to Hubble constant
$H_0$$\left(t_{\lim } =2H_{0}^{-1} \right)$. By this approach the
``acceleration effect'' arises in two limiting cases: on the light
cone and far off the cone (nonrelativistic local approximation).

In the first case we have nothing else but acceleration mimicry of a
kind.  The original cause for the observed effect (anomalous
frequency shift) is a conformal time inhomogeneity characterized
in the approximation                  ${\raise0.7ex\hbox{$ t
$}\!\mathord{\left/{\vphantom{t t_{\lim }
}}\right.\kern-\nulldelimiterspace}\!\lower0.7ex\hbox{$ t_{\lim }
$}} <<1$ by a uniform time acceleration [1,2] that is equal to
$\frac{d^{2} t'}{dt^{2} } =H_{0} $ (corresponding mimiced
acceleration $W=c\frac{d^{2} t'}{dt^{2} } =cH_{0} $).

In the second case $w=\frac{c}{2} H_{0} $ characterizes an
accelerated motion of the observer's local co-moving (noninertial!)
reference system. In noninertial RS itself this acceleration is
opposite in sign, being background in character (i.e. it should be
recorded in every point).

This conclusion is also supported by elementary dynamic
considerations, within the scope of a single-particle Lagrangian
model for motion in noninertial RS. As demonstrated in [23],
determining the conformal-invariant action for a free large-mass
particle in the standard way, one can have the following expression
for the single-particle Lagrangian

\begin{equation} \label{GrindEQ__5_2_}
L\left(\mathop{r}\limits_{\_ } ,\mathop{v}\limits_{\_ }
,t\right)=-mc^{2} \left(1-\frac{v^{2} }{c^{2} } \right)^{\frac{1}{2}
} \left\{\left(1-\frac{r}{2r_{0} } \right)^{2}
-\left(\frac{ct}{2r_{0} } \right)^{2} \right\}^{-1} ,
\end{equation}
from where, neglecting the terms $\frac{v}{c} ,\; \frac{r}{2r_{0} }
\; ,and\; \frac{{\rm ct}}{{\rm 2r}_{{\rm 0}} } $ with respect to
unity, one can obtain

\begin{equation} \label{GrindEQ__5_3_}
L_{0} \left(\mathop{r}\limits_{\_ } ,\mathop{v}\limits_{\_ }
\right)=\frac{m\mathop{v}\limits_{\_ } ^{2} }{2} -\frac{mc^{2}
}{r_{0} } r.
\end{equation}
This expression complies with the Lagrangian describing a
nonrelativistic motion of a large-mass particle under the action of
the inertial force  $\mathop{f}\limits_{\_ } $ given in the form

\begin{equation} \label{GrindEQ__5_4_}
{\bf f}=-grad\left(\frac{mc^{2} }{r_{0} } r\right)=-\frac{mc^{2}
}{r_{0} } \frac{{\bf r}}{r} =-\frac{mcH_{0} }{2} \frac{{\bf r}}{r} .
\end{equation}
As the reference point is not assigned, the distance r may be set
from an arbitrary chosen point in RS under consideration, the
situation is completely in line with an idea of the presence of a
negative background acceleration determined by the spatial expansion
rate in accordance with$w=\frac{1}{2} cH_{0} $.

It is seen that, within the approach proposed, the experimentally
recorded manifestations of the cosmological expansion are caused by
a noninertial character of the observer's reference system.
Actually, in this case there is no gravity. But the effect
associated with the presence of the universal background
acceleration may be interpreted proceeding from the equivalence
principle in terms of the existing stationary background
gravitational field. The more so that a negative sign of the
acceleration is indicative of the decreased expansion rate in
accordance with the standard dynamic pattern of the effect exerted
by gravity on the behavior of the normal matter after the Big Bang.

Of particular importance is the fact that the mimic acceleration W
computed from the observation data (i.e. based on experiments with
propagation of electromagnetic signals) is exactly two times as
great as the ``mechanic'' acceleration w directly determined from
the conformal transformations in Galilean-Newtonian approximation
(i.e. in the geometrically ``flat'' limit).

In a curious way, this situation is similar to the case of the light
beam deflection by the gravitating center. As known, neglect of the
space-time curvature by Einstein in 1911 (see [24]) has lead to the
result half as great as that obtained by him four years later within
the scope of GR (see [25]), to be in accord with the experimental
data.  Also, it should be noted that, conceptually and numerically,
values of the acceleration w are extraordinary close to the value of
the Milgrom \textit{minimal acceleration} appearing as a fundamental
parameter in the MOND (Modified Newton Dynamics, [26,27]) model that
has been proposed as an alternative for the cold dark matter
concept.

And, finally, note that the ``equivalent'' stationary background
gravitational field with the intensity $\sim c H_0$ present at every
point of the Metagalaxy, in principle, may play a role of the
alternative dark energy to do away with the present-day deficiency
of the gravitating matter density in the Universe. Indeed, if we
make use of a ``naive'' static field pattern and assume that the
energy density of a background gravitational field may be found as
$\rho _{F} =\alpha \left(\frac{cH_{0} }{G} \right)^{2} $, where
$\alpha $ -- numerical parameter, a perfectly acceptable result,
$\rho _{F} =\frac{2}{3} \rho _{c} $, where $\rho _{c} =\frac{3}{8\pi
} \frac{c^{2} H_{0}^{2} }{G} $ -- critical density, is achieved at
$\alpha =\frac{1}{4\pi } $. It is understood that such
considerations and estimates are not rigorous. A relativistically
consistent description of noninertial reference systems necessitates
the use of the GR techniques.

\section{Concluding remarks.}
\setcounter{equation}{0}

The principal outcome of this paper is the demonstration that the
observed features of the modern Universe expansion stage may be
described consistently within the scope of the space-time conformal
geometry, without the direct use of a pseudo-Riemann geometrical
model of GR. Non-Euclidean character of the space-time manifold is
presented in the proposed approach by conformal time inhomogeneity.
Characteristically, the experimentally recorded local expansion
effects in this case have to be interpreted as a manifestation of
the noninertiality of observer's reference system (``expanding''
RS), in every point of which there is a background acceleration
directed to the observation  point, its numerical value being
determined by the expansion rate (or by the observed value of Hubble
parameter).

It is clear that such a theoretical model looks practically like a
phenomenological one. But with this model there is a possibility not
only to simulate the observed effects but also to have the
experimentally checkable predictions. Moreover, even the fact that
within this model (i. e. without resort to GR) an analytical
dependence of the location distance on the red shift has been
established and an explicit expression for the cosmological
expansion law has been derived is undoubtedly of special interest.

Also, it is inferred that the necessary theoretical basis for
description of the processes associated with nonstationarity of the
spatial scales has been actually developed before the creation of GR
-- by 1911. In 1911,  in principle, it could have been used to
predict (later experimentally revealed) approximate linear
dependence between the recession rate and the corresponding distance
(Hubble law),  intermediate maximum at the point $z\approx 0.5$
(accelerated Universe expansion), and anomalous frequency shift
appearing in the location-type experiments (Pioneer anomaly).

It goes without saying that, due to a pure kinematic character of
the approach, gravitational sources as such are lacking in the
proposed expansion pattern. Nevertheless, they may be inferred on
the basis of the Einstein equivalence principle as a background
gravitational field, whose ``intensity'' as well as the associated
acceleration is determined by the Hubble constant $H_0$.

In the proposed model the Hubble constant $H_0$ acts as a parameter
that, in combination with the speed of light c, determines the
space-time dimensions for the observable region of the Universe. In
this case the constants c and $H_0$ are significantly differing in
their status. The presence of the upper limit for speed is of a
fundamental character, being organically involved in the SR bases.
However, up to the present no arguments in favor of the existing
constraints on the numerical value of Hubble parameter as inferred
from a certain general physical principle have been put forward. By
author's opinion, the Maximum Tension Principle proposed by Gibbons
in 2002 [28] looks very promising in this respect. By this
principle, in nature there is an upper limit for the rate of a
change in the energy-momentum (\textit{maximum force}
$\left(\frac{dp}{dt} \right)_{\lim } $and \textit{maximum power}
$\left(\frac{dE}{dt} \right)_{\lim } $), and the corresponding
numerical values of these limiting parameters are related to
Einstein's gravitational constant ${\raise0.7ex\hbox{$ G
$}\!\mathord{\left/{\vphantom{G c^{4}
}}\right.\kern-\nulldelimiterspace}\!\lower0.7ex\hbox{$ c^{4}  $}} $
as follows: $\left(\frac{dp}{dt} \right)_{\lim } =\frac{c^{4} }{4G}
$, $\left(\frac{dE}{dt} \right)_{\lim } =\frac{c^{5} }{4G} $.

Of course, in the absence of a developed theoretical concept based
on the principle, only heuristic considerations and approximate
quantitative assessments are possible. But such assessments lead to
the results of particular interest. To illustrate, let us consider
an expression for the acceleration determined as a ratio of maximal
force $F_{\max } =\frac{c^{4} }{4G} $ to the Universe mass, the
latter being determined as $M_{u} =\rho _{u} V_{u} $, where $\rho
_{u} $ -- average density of the Universe mass, $V_{u} $ -- its
volume. Assuming that $\rho _{u} $ is coincident with the critical
density$\rho _{c} =\frac{3}{8\pi } \frac{H_{0}^{2} }{G} $, whereas
the volume is determined from $V_{u} =\frac{4\pi }{3} R_{u}^{3} $,
where $R_{u} =cH_{0}^{-1} $, we can derive for  $M_{u} $ the
following expression: $M_{u} =\frac{1}{2} \frac{c^{3} }{GH_{0} } $.
As a result, for the sought quantity we have$w=\frac{F_{\max }
}{M_{u} } =\frac{1}{2} cH_{0} $, that is surprisingly coincident
with the background acceleration quantity found by us. Thus, $H_0$
observed presently looks like a limiting value (or close to the
limit), the limit, due to a fundamental character of the constant
${\raise0.7ex\hbox{$ c^{4} $}\!\mathord{\left/{\vphantom{c^{4}
G}}\right.\kern-\nulldelimiterspace}\!\lower0.7ex\hbox{$ G $}} $,
being determined exclusively by the matter amount in the Metagalaxy.

Even though the coincidences of this kind should not be
overestimated, it is not improbable that just a consistent allowance
for the constraints imposed by the Maximum Tension Principle sets
forth the way to possible modification of the traditional GR for
description of the gravity in  extreme conditions.

\end{document}